\documentclass[sigconf,authorversion,nonacm]{acmart} 

\acmSubmissionID{921}

\usepackage{amsfonts}
\usepackage{bm} 
\usepackage{subfigure}
\usepackage{multirow}
\usepackage{multicol}
\usepackage{wrapfig}
\usepackage{sidecap}
\usepackage{rotating}
\usepackage{pifont}
\usepackage{graphicx}

\newcommand{\myeqref}[1]{Eq.~\ref{#1}}

\AtBeginDocument{%
  \providecommand\BibTeX{{%
    \normalfont B\kern-0.5em{\scshape i\kern-0.25em b}\kern-0.8em\TeX}}}

\setcopyright{acmcopyright}
\copyrightyear{2023}
\acmYear{2023}
\acmDOI{XXXXXXX.XXXXXXX}

\sidecaptionvpos{figure}{c}

\begin{document}

\title{DiffXPBD : Differentiable Position-Based Simulation of Compliant Constraint Dynamics}

\author{Tuur Stuyck}\affiliation{
  \institution{Meta Reality Labs Research}
  \country{USA}
}

\author{Hsiao-yu Chen}\affiliation{
  \institution{Meta Reality Labs Research}
  \country{USA}
}

\renewcommand{\shortauthors}{Stuyck and Chen}

\begin{abstract}
We present DiffXPBD, a novel and efficient analytical formulation for the differentiable position-based simulation of compliant constrained dynamics (XPBD). 
Our proposed method allows computation of gradients of numerous parameters with respect to a goal function simultaneously leveraging a performant simulation model. The method is efficient, thus enabling differentiable simulations of high resolution geometries and degrees of freedom (DoFs). Collisions are naturally included in the framework.
Our differentiable model allows a user to easily add additional optimization variables. Every control variable gradient requires the computation of only a few partial derivatives which can be computed using automatic differentiation code. We demonstrate the efficacy of the method with examples such as elastic cloth and volumetric material parameter estimation, initial value optimization, optimizing for underlying body shape and pose by only observing the clothing, and optimizing a time-varying external force sequence to match sparse keyframe shapes at specific times. Our approach demonstrates excellent efficiency and we demonstrate this on high resolution meshes with optimizations involving over 26 million degrees of freedom. Making an existing solver differentiable requires only a few modifications and the model is compatible with both modern CPU and GPU multi-core hardware.  
\end{abstract}



\begin{teaserfigure}
  \includegraphics[width=\textwidth]{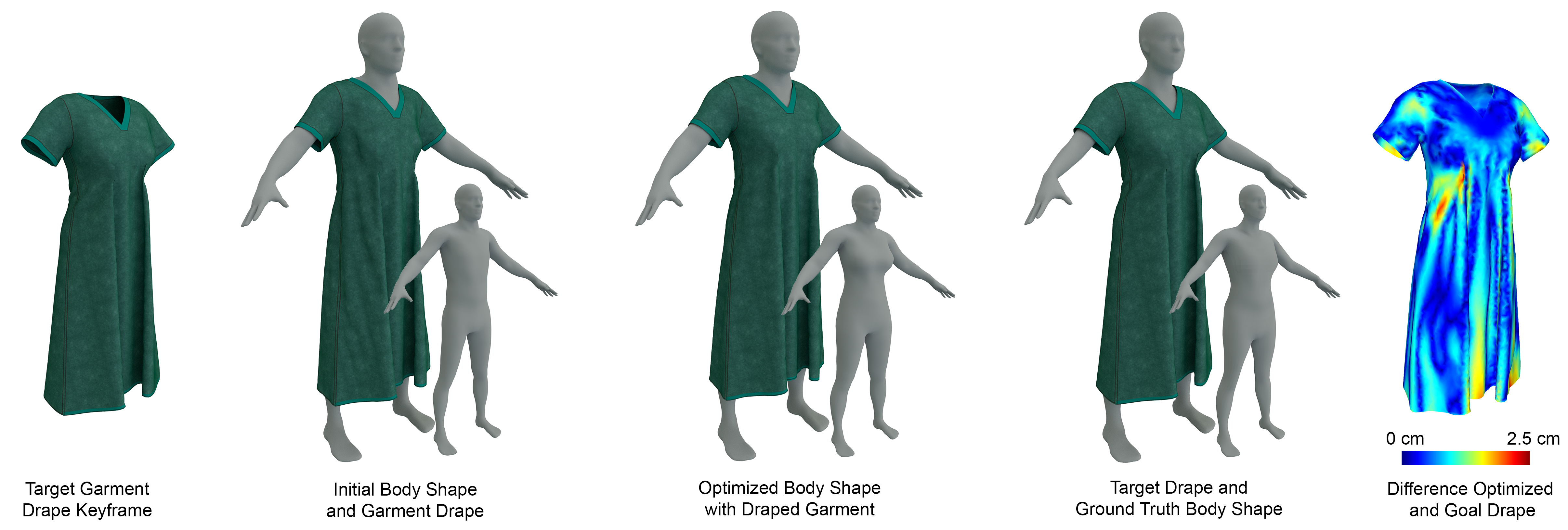}
  \caption{We present a fully analytically differentiable solver that allows to obtain gradients with respect to a desired parameter in order to minimize a goal function. In this example, we show how coefficients for the shape of a statistical body model can efficiently be computed in order to minimize the distance of the draped clothing to an observed reference. Our differentiable solver can differentiate through collisions of the cloth with the underlying body shape which allows the gradient information to propagate. Left shows the target drape. Next, we show the original body shape with the draped cloth. The middle figure shows the final optimized body shape that produces a drape close to the goal shape. The goal drape and ground truth body shape are shown second to the right. The rightmost figure visualizes the distance of the estimated garment drape to the ground truth drape. Note how they closely coincide where the garment touches the body, indicating a successful optimization result.}
  \label{fig:teaser}
\end{teaserfigure}

\maketitle

\section{Introduction}

Differentiable simulation enables integration of physics-based models with data-driven methods in a seamless and efficient way. By making the simulation model differentiable, we can compute gradients of the output variables with respect to the model parameters, which can be optimized through gradient-based methods. This opens up a range of possibilities for tasks such as model identification, material estimation, inverse design, where we want to fit the simulation model to data or optimize certain parameters to achieve specific goals. Differentiable simulation can also enable us to create more realistic and accurate simulations by incorporating real-world data into the simulation process. Recent advances in capture systems \cite{halimi2022garment, chen2021capturing, White2007CAO, Wang_2022_CVPR, Wang_2023_CVPR} provide high quality data that can be used to learn the simulation parameters and fine-tune the simulation model to better match real-world behavior. Physics-based simulation methods have shown widespread success and many different techniques have been proposed. The seminal work of \citet{baraff1998large} introduced an implicit integration scheme enabling simulations which remain stable for large time steps, resulting in efficient simulations. This method is still commonly used in top-tier animation studios \cite{kim2020dynamic}. Since then, many novel approaches have been proposed to tackle different shortcomings. The introduction of Position-Based Dynamics (PBD) \cite{muller2007position} enabled a unified high performance solver that maps efficiently to modern parallel hardware such as multi-core CPUs and GPUs. Follow up work introduced eXtended Position-Based Dynamics (XPBD) \cite{macklin2016xpbd} which resolves the iteration and time step dependent stiffness issue of PBD. Projective Dynamics (PD) \cite{bouaziz2014projective} introduced a fast local-global solver for implicit time integration of FEM simulations with elastic energies in a quadratic form. \citet{stuyck2018cloth} provides an overview of several simulation techniques.

Several of these methods have been extended to be differentiable for dynamic simulations and we are the first to present an analytical differentiable formulation for the XPBD simulation framework. The differentiable solver can easily be extended to novel constraints by adding a few partial derivatives per constraint. We empirically show that our method scales to high simulation resolutions and DoFs with respect to prior work. Our contributions are the following :
\begin{itemize}
    \item We present a differentiable formulation for position-based simulation of compliant constraint dynamics. The method is efficient, scales to large number of DoFs, and maintains all advantages of the forward simulation model enabling parallel cpu and gpu implementations.
    \item Our method naturally differentiates through the constraints formulation which seamlessly enables differentiable simulation of different potentially coupled phenomena, including self-collisions and collisions with the environment.
    \item The approach requires only a small addition to the forward simulation model and as such, it can be easily added to existing solvers. The backward solve involves sparse matrices and can be implemented using off-the-shelf libraries.
\end{itemize}
We show efficient computation of derivatives of high DoFs and high geometric resolutions for a variety of tasks.

\section{Related Work}

Differentiable simulation has been applied in recent research \cite{liang2019differentiable, qiao2020scalable} to system identification and for inferring material parameters from observations \cite{strecke2021_diffsdfsim, hu2019chainqueen}. It has applications in computer graphics, vision, robotics and many others. Several of these techniques rely on the adjoint method for gradient computation. Over the last decades, the adjoint method has been successfully applied to differentiate various dynamic simulation models. This includes fluid simulation \cite{mcnamara2004fluid} and implicit simulation of cloth dynamics \cite{wojtan2006keyframe}. The adjoint method continues to be successfully applied with a focus on time-dependent deformation problems with contact \cite{gjoka2022differentiable}. Notoriously, cloth simulations frequently undergo numerous contacts. Because of this, work has focused specifically on the handling of differentiable simulation in contact heavy scenarios \cite{zhong2022differentiable}. \citet{liang2019differentiable} introduced a differentiable approach for handling cloth collisions using a linear complementary problem. \citet{du2021_diffpd} introduced a differentiable approach to Projective Dynamics using the adjoint method, enabling fast differentiable simulations. Follow up work extends this research to enable  cloth simulations with dry frictional contact \cite{diffcloth}. Others focus on the optimization of static cloth simulations \cite{bartle2016physics, umetani2011sensitive}. \citet{DiffSimCourse} provide an overview of differentiable simulation methods.

Differentiable soft body simulation models \cite{geilinger2020add, hahn2019real2sim} and estimating material properties from scanned volumetric objects using differentiable simulation as an inverse design problem has been active domain of research \cite{weiss2020correspondence}. \citet{veo} introduced differentiable point-based simulation for material estimation of soft deformable bodies coupled with neural radiance field representations \cite{mildenhall2021nerf}. Similarly, parameter estimation for cloth simulations has been a focus of attention \cite{Larionov, wang2011data, miguel12}. \citet{guo2021inverse} showed that is possible to estimate body shape and pose given a point cloud by differentiating through clothing simulations. A full end-to-end system is presented by~\citet{gradsim} where they propose a system identification technique by combining differentiable simulation with differentiable rendering which allows them to backpropagate gradient information from pixels in a video sequence. This pipeline enables them to optimize for control variables to reproduce image observations directly. Other simulation models like the Material Point Method have successfully been made differentiable \cite{hu2019chainqueen} and have been used for a variety of control tasks \cite{hu2019difftaichi}.

In addition to analytical formulations of these differentiable simulation models, specialized differentiable programming frameworks focusing on physics-based simulation applications are becoming more commonplace \cite{warp2022, taichi}. The research field has made great progress enabling dynamic differentiable simulation for numerous applications. However, several issues still remain. Performance and memory usage is a primary concern for any differentiable method. With our efficient and highly parallelizable formulation, we demonstrate optimizations involving high resolution geometries and DoFs.

\section{Background}

\subsection{XPBD: Position-Based Simulation of Compliant Constrained Dynamics}

XPBD is an efficient unified simulation model, capable of producing real-time simulations. The constraint-based formulation can be parallelized \cite{fratarcangeli2016vivace} and implemented on modern GPU and multi-core CPU hardware. As a result, the model showcases much better performance compared to expensive non-linear solvers. The method solves Newton's equations of motion given by $\mathbf{M}\ddot{\mathbf{x}} = -\nabla U^{\top}(\mathbf{x}),$ where $\mathbf{x}$
are the $V$ vertex positions and $\mathbf{M}$ is the mass matrix.
The energy potential $U(\mathbf{x})$ is formulated in terms of a vector of constraint functions $\mathbf{C} = \left[C_1(\mathbf{x}), \cdot \cdot \cdot, C_m(\mathbf{x})\right]^{\top}$ and inverse compliance matrix \(\bm{\alpha}^{-1}\) as
\begin{equation}
U(\mathbf{x}) = \frac{1}{2} \mathbf{C}(\mathbf{x})^{\top} \bm{\alpha}^{-1} \mathbf{C}(\mathbf{x})    
    \label{eq:xpbdUenergy}
\end{equation}
Implicit Euler time integration results in the constraint multiplier updates $\Delta \bm{\lambda}$ at iteration $i$ being computed as 
\begin{align}
    (\nabla \mathbf{C}(\mathbf{x}_i)^{\top}\mathbf{M}^{-1}\nabla\mathbf{C}(\mathbf{x}_i) + \tilde{\bm{\alpha}}) \Delta \bm{\lambda} = - \mathbf{C}(\mathbf{\mathbf{x}_i}) - \tilde{\bm{\alpha}}\bm{\lambda}_i \label{eq:xpbd}
\end{align}
where $\tilde{\bm{\alpha}} = \bm{\alpha} / \Delta t^2$. Given $\Delta \bm{\lambda}$, the position update is computed as
\begin{align}
\Delta \mathbf{x} = \mathbf{M}^{-1} \nabla \mathbf{C}(\mathbf{x}_i)\Delta \bm{\lambda}
\end{align}
With external forces $\mathbf{f}_{\text{ext}}$ acting on the system. The state $\mathbf{q}_n = \left(\mathbf{x}_n,\mathbf{v}_n\right)$ at time step $n$ consisting of positions $\mathbf{x} \in \mathbb{R}^{3V}$ and velocities $\mathbf{v} \in \mathbb{R}^{3V}$ is updated as
\begin{equation}
\begin{aligned}
 \mathbf{x}_{n+1} 
 &= \mathbf{x}_{n} + \Delta \mathbf{x}\left(\mathbf{x}_{n+1} \right) + \Delta t \left( \mathbf{v}_n + \Delta t \mathbf{M}^{-1} \mathbf{f}_{\text{ext}} \right) \\
 \mathbf{v}_{n+1} 
 &= \frac{1}{\Delta t} \left( \mathbf{x}_{n+1} - \mathbf{x}_{n} \right)
\end{aligned}
\label{eq:xpbdUpdateRule}
\end{equation}

\subsection{The Adjoint Method}

The gradients $d \phi / d \mathbf{u}$ required to minimize a goal function $\phi$ with respect to control variables $\mathbf{u}$ are typically intractable to compute directly for dynamic simulations. The adjoint method provides an efficient solution to compute these derivatives in a single pass through the computation graph, regardless of the number of parameters. The gradients of the control variables with respect to some goal function can be computed as
\begin{equation}
\frac{d \phi}{d \mathbf{u}} = \frac{\partial \phi}{\partial \mathbf{Q}}\frac{d \mathbf{Q}}{d \mathbf{u}} + \frac{\partial \phi}{\partial \mathbf{u}}
\end{equation}
where $\mathbf{Q}$ represents the full set of states $\mathbf{q}_n$ across all times $N$. The adjoint method turns this intractable computation into a more efficient formulation by replacing the vector-matrix product containing $d\mathbf{Q} / d\mathbf{u}$ with an equivalent computation involving the adjoint of $\mathbf{Q}$, denoted by $\hat{\mathbf{Q}}$ which contains all adjoint states $\hat{\mathbf{q}}_n = \left(\hat{\mathbf{x}}_n \in \mathbb{R}^{3V},\hat{\mathbf{v}}_n \in \mathbb{R}^{3V}\right)$ over all times $N$. 
We refer to the supplemental material and \citet{wojtan2006keyframe} for an in-depth overview. The simulation moves the states forward in time using $\mathbf{q}_{n+1} = \mathbf{F}_n\left(\mathbf{q}_{n+1}, \mathbf{q}_n, \mathbf{u}\right)$, see \myeqref{eq:xpbdUpdateRule}. The adjoint states $\hat{\mathbf{Q}}$ are computed in a backward pass using \begin{equation}
    \hat{\mathbf{q}}_{n-1} =  \left( \frac{\partial \mathbf{F}_{n-1}}{\partial \mathbf{q}_n}  \right)^\top \hat{\mathbf{q}}_{n-1} + \left( \frac{\partial \mathbf{F}_n}{\partial \mathbf{q}_n} \right)^\top \hat{\mathbf{q}}_n + \left( \frac{\partial \phi}{\partial \mathbf{q}_n} \right)^{\top}
    \label{eq:adjointRule}
\end{equation}
after which the full derivative $\frac{d \phi}{d \mathbf{u}}$ is obtained using 
\begin{equation}
    \frac{d \phi}{d \mathbf{u}} = \hat{\mathbf{Q}}^{\top} \frac{\partial \mathbf{F}}{\partial \mathbf{u}} + \frac{\partial \phi}{\partial \mathbf{u}}
    \label{eq:gradientRule}
    \end{equation}

\section{Method}

\begin{figure}
    \centering
    \includegraphics[width=0.49\textwidth, trim={0 3cm 0 0},clip]{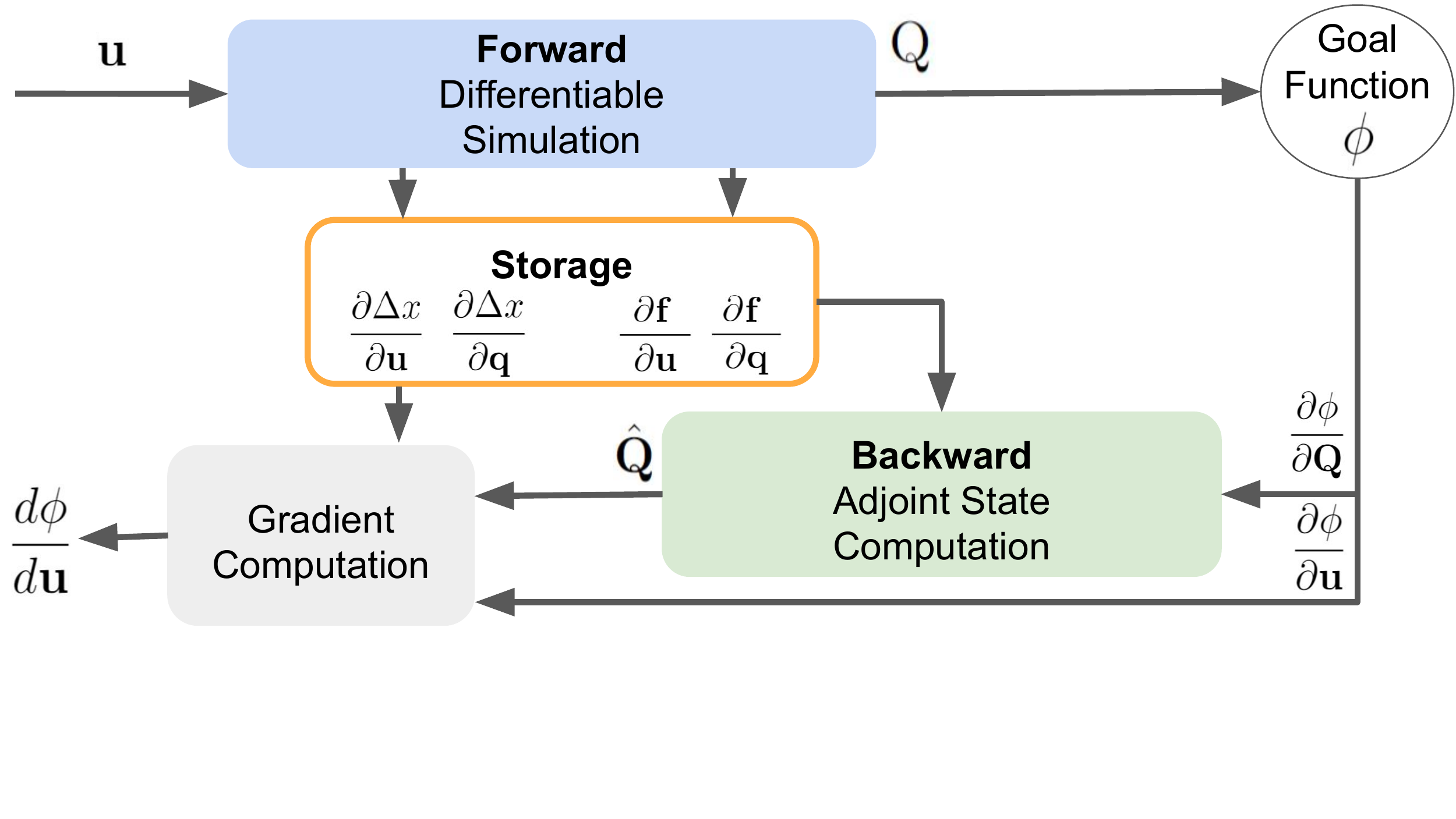}
    \caption{\emph{Overview of the different components of the gradient computation process.} The forward simulation computes and stores the quantities required in the backward pass. After which, the goal function gets evaluated which in turn starts the backward adjoint state computation pass. The total gradient $d \phi / d \mathbf{u}$ is obtained using~\myeqref{eq:gradientRule}.}
    \label{fig:overview}
\end{figure}

\subsection{Goal Function}

An overview of the method is illustrated in Figure~\ref{fig:overview}. We minimize a goal function $\phi$ by modifying the control variables $\mathbf{u}$. This function can be easily provided by the user and adapted based on the problem at hand. The most straight-forward metric is to compare simulation state $\mathbf{Q}$ directly with a given reference $\mathbf{Q^{*}}$ using 
\begin{align}
\phi(\mathbf{u}, \mathbf{Q}) = \frac{1}{2} \sum^N_{n=0} \left( ||\mathbf{W}_n(\mathbf{q}_n - \mathbf{q}^*_n) ||^2 + \beta || \mathbf{u}_n ||^2 \right)
\end{align}
where an optional regularization term with weight $\beta$ based on the control variable $\mathbf{u}$ has been added. The weight matrix $\mathbf{W}_n$ can be used to introduce relative importance. The goal function can be defined over all steps $N$ or a subset thereof. Any goal function or combination of goal functions can be used as long as the required derivatives $\partial \phi / \partial \mathbf{q}$ and $\partial \phi / \partial \mathbf{u}$ can be computed. Section~\ref{sec:bodyShapeOptimization} shows the use of a point-cloud-based goal function.

\subsection{Adjoint State Computation}
The adjoint evolution for the XPBD integration scheme is found by combining \myeqref{eq:xpbdUpdateRule} and \myeqref{eq:adjointRule}, see the supplemental material for the full derivation. We find
\begin{equation}
\begin{aligned}
    \hat{\mathbf{x}}_{n} 
    &= \hat{\mathbf{x}}_{n+1} + \left(\frac{\partial \Delta \mathbf{x}}{\partial \mathbf{x}}
    + \Delta t^2 \mathbf{M}^{-1} \frac{\partial \mathbf{f}_{\text{ext}}}{\partial \mathbf{x}} \right)^\top \hat{\mathbf{x}}_{n}
    + \frac{\hat{\mathbf{v}}_n}{\Delta t}
    - \frac{\hat{\mathbf{v}}_{n+1}}{\Delta t}
    + \frac{\partial \phi}{\partial \mathbf{x}}^\top \\
    \hat{\mathbf{v}}_n 
    &= \left( \frac{\partial \Delta \mathbf{x}}{\partial \mathbf{v}} + \Delta t^2 \mathbf{M}^{-1} \frac{\partial \mathbf{f}_{\text{ext}}}{\partial \mathbf{v}} \right)^\top \hat{\mathbf{x}}_{n} 
    + \Delta t \hat{\mathbf{x}}_{n+1} + \frac{\partial \phi}{\partial \mathbf{v}}^\top
\end{aligned}
\label{eq:rawAdjoint}
\end{equation}

After re-arranging and by substituting \(\hat{\mathbf{v}}_n\) we find

\begin{equation}
\begin{aligned}
&\left( I - \frac{\partial \Delta \mathbf{x}}{\partial \mathbf{x}} - \Delta t^2 \mathbf{M}^{-1} \frac{\partial \mathbf{f}_{\text{ext}}}{\partial \mathbf{x}} - \frac{1}{\Delta t}\frac{\partial \Delta \mathbf{x}}{\partial \mathbf{v}} - \Delta t \mathbf{M}^{-1}\frac{\partial \mathbf{f}_{\text{ext}}}{\partial \mathbf{v}} \right)^\top \hat{\mathbf{x}}_n \\ 
&= 2\hat{\mathbf{x}}_{n+1} - \frac{\hat{\mathbf{v}}_{n+1}}{\Delta t}  + \frac{\partial \phi}{\partial \mathbf{x}}^\top + \frac{1}{\Delta t} \frac{\partial \phi}{\partial \mathbf{v}}^\top
\end{aligned}
\label{eq:adjointXPBD}
\end{equation}
Assuming that the external forces are generally independent of the positions and velocities, and since $\partial \Delta \mathbf{x} / \partial \mathbf{v}$ evaluates to $\mathbf{0}$, we left multiply with the mass matrix $\mathbf{M}$. We solve this sparse symmetric linear system using Conjugate Gradients (CG) to obtain $\hat{\mathbf{x}}_n$. We then find $\hat{\mathbf{v}}_n$ using \myeqref{eq:rawAdjoint}.

\subsubsection{Linear System Filtering} Vertices can be fully constrained in the simulation due to Dirichlet boundary conditions. These pinned vertices are modeled using an infinite mass. To handle this robustly, we rely on modified Conjugate Gradient with pre-filtering \cite{baraff1998large} as presented by~\citet{tamstorf2015smoothed}.

\subsection{Position Update Derivative Computation}
\label{subsec:deltaxx}
For every constraint in the system, \myeqref{eq:adjointXPBD} requires us to differentiate through the position update in order to obtain $\partial \Delta \mathbf{x}/ \partial \mathbf{x}$. These derivatives can be computed analytically or using automatic or symbolic differentiation techniques. The derivatives can be accumulated in parallel in the same iterative fashion as the position updates, which allows the method to remain highly parallelizable. Analytically, these derivatives are computed as
\begin{equation}
\begin{aligned}
\frac{\partial \Delta \mathbf{x}}{\partial \mathbf{x}} &= \mathbf{M}^{-1} \left( \frac{\partial \nabla \mathbf{C}}{\partial \mathbf{x}} \Delta \bm{\lambda} + \nabla \mathbf{C} \frac{\partial \Delta \bm{\lambda}}{\partial \mathbf{x}} \right)
\end{aligned}
\end{equation}
Where we can omit the $\mathbf{M}^{-1}$ term since this vanishes when multiplied by the mass matrix in \myeqref{eq:adjointXPBD}. To keep the notation compact, we define $\nabla \mathbf{C}(\mathbf{x}_i)^{\top}\mathbf{M}^{-1}\nabla\mathbf{C}(\mathbf{x}_i) + \tilde{\bm{\alpha}} = \mathbf{J}$, and we group the terms in \myeqref{eq:xpbd} as $\mathbf{J} \Delta \bm{\lambda} = \mathbf{b}$. We can then compute the required quantities as
\begin{equation}
\begin{aligned}
\frac{\partial \Delta \bm{\lambda}}{\partial \mathbf{x}} 
\hspace{0.2cm} = -\mathbf{J}^{-1} \frac{\partial \mathbf{J}}{\partial \mathbf{x}} \mathbf{J}^{-1} \mathbf{b} + \mathbf{J}^{-1}\frac{\partial \mathbf{b}}{\partial \mathbf{x}} 
\hspace{0.2cm} = -\mathbf{J}^{-1} \left(\frac{\partial \mathbf{J}}{\partial \mathbf{x}} \Delta \bm{\lambda} - \frac{\partial \mathbf{b}}{\partial \mathbf{x}}\right)           
\end{aligned}
\end{equation}
where $\frac{\partial \mathbf{b}}{\partial \mathbf{x}}$ and $\frac{\partial \mathbf{J}}{\partial \mathbf{x}}\Delta \bm{\lambda}$ are computed as
\begin{equation}
\begin{aligned}
\frac{\partial \mathbf{b}}{\partial \mathbf{x}} 
\hspace{0.2cm} = -\frac{\partial \mathbf{C}}{\partial \mathbf{x}} - \frac{\partial \tilde{\bm{\alpha}}}{\partial \mathbf{x}}\bm{\lambda} - \tilde{\bm{\alpha}} \frac{\partial \bm{\lambda}}{\partial \mathbf{x}}
\hspace{0.2cm} = - \nabla \mathbf{C} - \tilde{\bm{\alpha}} \sum \frac{\partial \Delta \bm{\lambda}}{\partial \mathbf{x}}
\end{aligned}
\end{equation}
\begin{equation}
\begin{aligned}
\frac{\partial \mathbf{J}}{\partial \mathbf{x}}\Delta \bm{\lambda} 
&= \frac{\partial \nabla \mathbf{C}^T}{\partial \mathbf{x}} \mathbf{M}^{-1} \nabla \mathbf{C} \Delta \bm{\lambda} + \nabla \mathbf{C}^T \mathbf{M}^{-1} \frac{\partial \nabla \mathbf{C}}{\partial \mathbf{x}}\Delta \bm{\lambda} \\
&= \frac{\partial \nabla \mathbf{C}^T}{\partial \mathbf{x}} \Delta \mathbf{x} + \nabla \mathbf{C}^T \mathbf{M}^{-1} \frac{\partial \nabla \mathbf{C}}{\partial \mathbf{x}}\Delta \bm{\lambda}
\end{aligned}
\end{equation}

\subsubsection{Derivative Verification}

According to the XPBD formulation, we have the relation $\mathbf{M} \frac{\partial \Delta \mathbf{x}}{\partial \mathbf{x}} = \frac{\partial\mathbf{f}}{\partial \mathbf{x}}$, where $\mathbf{f} = - \nabla_x U^{\top}\left( \mathbf{x}_{n+1} \right)$, and $\frac{\partial\mathbf{f}}{\partial \mathbf{x}}$ is the Hessian of an elastic potential $U$. To validate the derivative, one can rely on the symmetry property of the Hessian and verify the symmetry of $\mathbf{M} \frac{\partial \Delta \mathbf{x}}{\partial \mathbf{x}}$. In addition, the internal forces inside each element should sum to zero, which implies that $\sum_i \frac{\partial\mathbf{f_i}}{\partial \mathbf{x}} = 0 $, for all vertices $i$ inside the element. This suggests that the rows of $\mathbf{M} \frac{\partial \Delta \mathbf{x}}{\partial \mathbf{x}}$ sum to zero, and by symmetry the column should sum to zero as well. A more detailed discussion can be found in the documents by ~\citet{kim2020dynamic}.

\subsubsection{Positive Definite Projection}
Although a direct solver can be used to solve \myeqref{eq:adjointXPBD}, iterative solvers like Preconditioned Conjugate Gradients (PCG) are preferred due to its efficiency. PCG requires the matrix to be semi-positive-definite. Therefore, we project all derivative blocks to positive definiteness \cite{nocedal}. The projection can be computed in parallel over all derivatives and happens only once at the end of every step.

\subsection{Control Variable Derivative Computation}

For every control variable $\mathbf{u}$ we wish to obtain a gradient for, we need to evaluate \myeqref{eq:gradientRule} which requires $\partial \mathbf{F} / \partial \mathbf{u}$. For control variables influencing the position updates, we need to compute and store $\partial \Delta \mathbf{x} / \partial \mathbf{u}$. These derivatives can be computed using a similar derivation as explained in Section~\ref{subsec:deltaxx}. For those influencing the external forces, additional $\partial \mathbf{f}_{\text{ext}} / \partial \mathbf{u}$ terms are needed. 

\subsection{Collision Handling}

In order to include collision handling in our differentiable framework, we model collisions as stiff springs using compliant constraints to maintain a closest separation distance of at least a user-set thickness between the garment layers as well as the body surface. The approach is compatible with any collision detection strategy such as continuous time collision detection.

\section{Example Applications}

Our method provides a way to efficiently compute gradients analytically with respect to any control variable. To illustrate this, we present several example applications leveraging our optimization pipeline. Our method is not limited to these specific examples.

\subsection{Cloth Material Parameter Estimation}

We use an orthotropic Saint Venant-Kirchhoff membrane energy model combined with a discrete bending \cite{bender17} term to model the fabric properties. Different material models are equally applicable and our proposed differentiation method is not limited to these. Per triangle with area \(A\), the membrane model has an inverse compliance matrix of the form
$$
	\bm{\alpha}_{\triangle}^{-1} = A\begin{bmatrix} C_{00} & C_{01} & \\ C_{01} & C_{11} & \\ & & C_{22} \end{bmatrix},
$$
where \(C_{ij}\) are the compliance coefficients. The constraint function for each triangle is then defined to be the Green strain $\epsilon$ in Voigt notation $\mathbf{C}_{\triangle}(\mathbf{x}) = (\epsilon_{uu}, \epsilon_{vv}, 2\epsilon_{uv})^{\top},$ where subscripts $u$ and $v$ indicate warp and weft directions, respectively.
For bending, the inverse compliance matrix is given by the scalar bending stiffness: $\bm{\alpha}_{\text{bend}}^{-1} = b$. We optimize directly over the compliance coefficients and bending parameter by choosing the parameter set to be
\begin{equation}
\bm{\gamma} := (C_{00}, C_{11}, C_{01}, C_{11}, b), \ \bm{\gamma} \in \bm{\Gamma}
\end{equation}
where the parameters are constrained to be in the feasible set $\bm{\Gamma}$. In order to optimize for these parameters, we compute and store $\partial \Delta \mathbf{x} / \partial \bm{\gamma}$ during the constraint solve at every step.

\subsection{Volumetric Material Parameter Estimation}
To model inhomogeneous elastic objects, we utilize the stable Neo-Hookean energy~\cite{smith2018}, following the XPBD formulation by~\citet{macklin2021}, with the hydrostatic constraint $\mathbf{C}_H = \textit{det}(\mathbf{F})-1$ to preserve volume, and the deviatoric constraint $\mathbf{C}_D = \sqrt{\textit{tr}(\mathbf{F}^T\mathbf{F})}$ to penalize stretching, where $\mathbf{F}$ is the deformation gradient. We assume that the object can be composed of different materials, characterized by sets of first and second Lam\'{e} parameters, denoted as $(\bm{\mu}, \bm{\lambda})$. Our optimization process aims to identify the set of parameters $(\mathbf{a}, \mathbf{b})$ such that $\mathbf{a}^2 = \bm{\mu}$ and $\mathbf{b}^2 = \bm{\lambda}$. By optimizing over $\mathbf{a}$ and $\mathbf{b}$, we ensure that the optimized values are within the physical range, as the Lam\'{e} parameters are strictly positive.

\subsection{Body Shape Optimization}

Given a statistical body model such as SMPL \cite{SMPL} but not limited to this specific model, we can estimate the PCA shape coefficients $\bm{\alpha}$ that would minimize a given goal function defined on the simulated cloth mesh. The connection between the cloth and the body is established through the collisions and gradients can flow through freely. We accumulate gradients with respect to the coefficients using the chain rule
\begin{align}
\frac{\partial \Delta \mathbf{x}_\text{cloth-body collision}}{\partial \bm{\alpha}} = 
\frac{\partial \Delta \mathbf{x}_\text{cloth-body collision}}{\partial \mathbf{x}_\text{body} } \frac{\partial \mathbf{x}_\text{body} }{\partial \bm{\alpha}}
\end{align}

\subsection{Skeleton Pose Optimization}
\label{subsecLposeTheory}

Similarly as for the body shape optimization, by applying the chain rule and differentiating through the linear blend skinning directly, we obtain the gradient with respect to the skeleton joint angles $\mathbf{\tau}$ using 
\begin{align}
\frac{\partial \Delta \mathbf{x}_\text{cloth-body collision}}{\partial \bm{\tau}} = \frac{\partial \Delta \mathbf{x}_\text{cloth-body collision}}{\partial \mathbf{x}_\text{body} } \frac{\partial \mathbf{x}_\text{body} }{\partial \bm{\tau}}
\end{align}

\subsection{Initial Value Optimization}
\label{subsecInitialValueOpt}

Following~\myeqref{eq:gradientRule}, the gradient with respect to the initial position and velocities can be directly obtained from the adjoint states without the need for computing additional control variable derivatives of the position updates.

\subsection{Keyframe Simulation}

The method can also be used for motion in-betweening or stitching distinct simulations together. Provided with a few keyframes, the method can automatically optimize for an external time-varying force sequence that pushes the clothing through the keyframes at the desired times. We optimize directly for all external forces per particle per step resulting in a high dimensional optimization problem. The external force gradients can be directly obtained from the adjoint velocities as they do not influence the position updates directly.

\section{Results}

We demonstrate the efficacy of our method with respect to several applications. All differentiable simulations are dynamic and accompanying videos can be found in the supplemental material. The resolutions of the simulated objects are reported in Table~\ref{tab:timingsForward}. All optimizations are minimized using gradient descent with fixed step size.

\subsection{Cloth Material Parameter Estimation}

\begin{figure}
  \centering
    \subfigure[Initial]{\includegraphics[width=0.48\columnwidth]{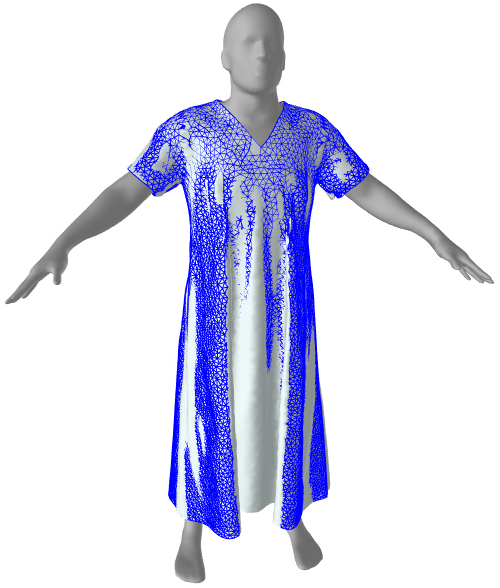}}
    \subfigure[Optimized]{\includegraphics[width=0.48\columnwidth]{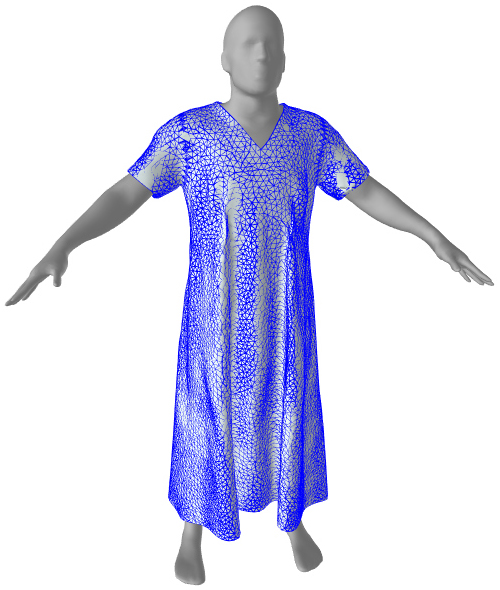}} 
    \caption{\emph{Bend Stiffness Optimization.} Material parameter estimation is effective even when the clothing is interacting with an underlying body. The optimized fabric coincides closely with the target shape shown in the blue wireframe.}
    \label{fig:bendonBodyOptimization}
\end{figure}

\begin{figure*}
    \centering
    \includegraphics[width=0.98\textwidth]{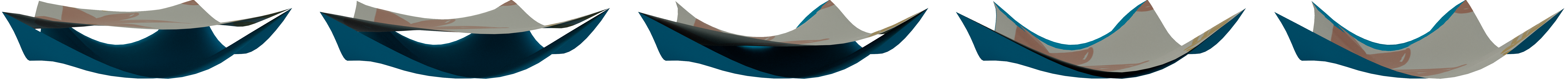}
    \caption{\emph{Young's Moduli Optimization.} We show different iterations of a cloth swatch draping under gravity with its corners pinned. The simulated cloth is textured and the goal shape is shown in blue. We optimize for the elastic material properties such that the simulated cloth reaches the desired pose at the requested frame. Initially, the material is too stiff and sags insufficiently under gravity to reach the target state. The optimization converges quickly to a looser material that matches the target at the specified frame.}
    \label{fig:stiffnessOptimization}
\end{figure*}

Our method is capable of optimizing the material parameter while the clothing is draped on a body for which the collisions are taken into account. We demonstrate before and after estimation of the bending stiffness for a dress draped on a static body in Fig.~\ref{fig:bendonBodyOptimization} where the keyframe is shown in blue. Similarly, we estimate in-plane elastic material properties. Figure~\ref{fig:stiffnessOptimization} demonstrates the optimization of the Young's moduli in both warp and weft direction such that a target shape gets matched at a specific frame in a dynamic simulation. In this example, all triangles in the swatch share the same material.

\subsection{Volumetric Material Parameter Estimation}
\begin{figure}
    \centering
    \includegraphics[width=0.64\linewidth]{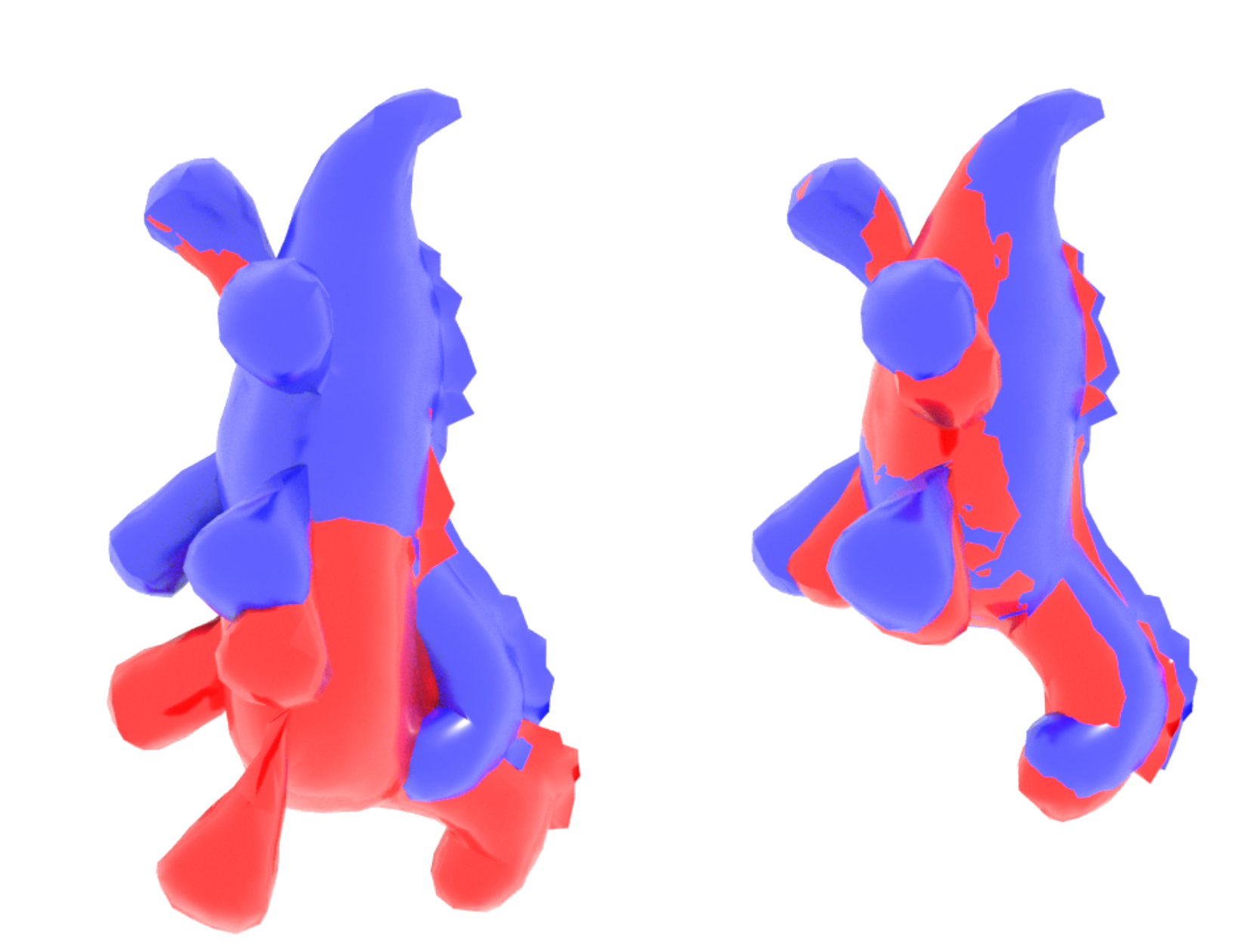}
    \caption{\emph{Volumetric Material Optimization.} Our method estimates the material of an inhomogenous elastic body (red) to fit a target (blue), starting from an initial guess (left) to an optimized result (right).}
    \label{fig:volumetricOptimization}
\end{figure}

We demonstrate estimation of the material parameters of an inhomogeneous volumetric elastic object. The target object is composed of two materials with different stiffnesses, and we create a target shape by fixing one end and allowing the rest to deform under gravity. We try to match this boundary shape by optimizing for 100 materials randomly distributed over all tetrahedrons, resulting in 200 DoFs. Fig.~\ref{fig:volumetricOptimization} demonstrates the effectiveness of our method in finding a set of material parameters that generates similar behavior to the target, even when the initial guess is vastly different. 

\subsection{Body Shape Optimization}
\label{sec:bodyShapeOptimization}
\begin{figure}
    \centering
    \includegraphics[width=0.4\textwidth]{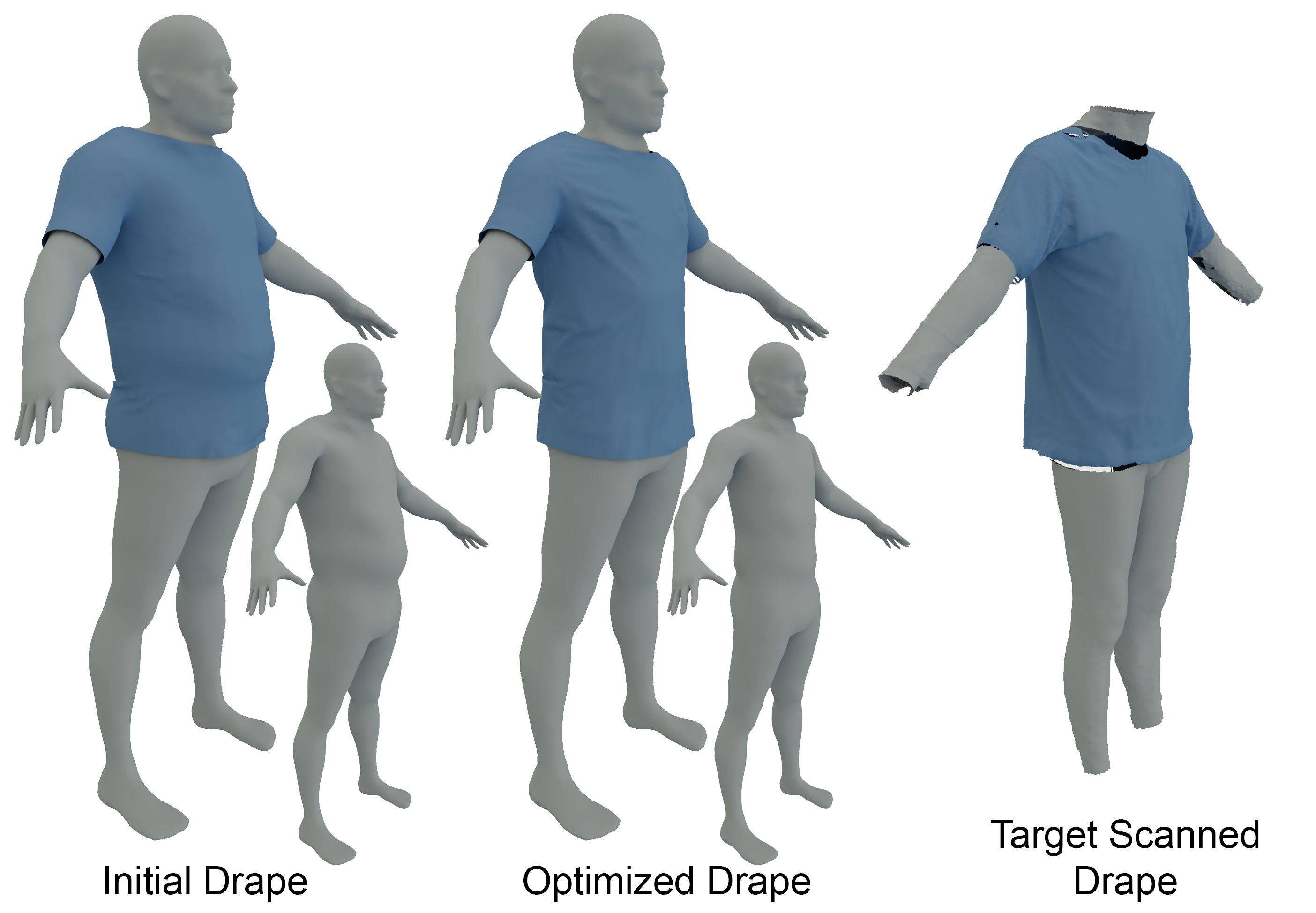}
    \caption{\emph{Body Shape Optimization From Scan.} We show successful body shape optimization given a high resolution scan.}
    \label{fig:bodyShapeOptimizationRealScan}
\end{figure}

We show optimization of the coefficients of a statistical body model so that a draped garment matches a given cloth observation as closely as possible. This is demonstrated in Figure~\ref{fig:teaser}. We optimize for all 2781 body PCA coefficients simultaneously. Fig.~\ref{fig:bodyShapeOptimizationRealScan} shows body shape optimization using a scan of a real person captured wearing a t-shirt. Since the scan has a different topology than the simulated garment, we use a goal function based on the closest distance of the simulated vertices to the scan where closest points computations are recomputed every iteration and the t-shirt geometry is segmented out of the scan. The segmented t-shirt scan contains 144,556 vertices and 288,034 faces and the optimized body shape contains 7,324 vertices and 14,644 faces.

\begin{figure}
    \centering
  \includegraphics[width=0.4\textwidth]{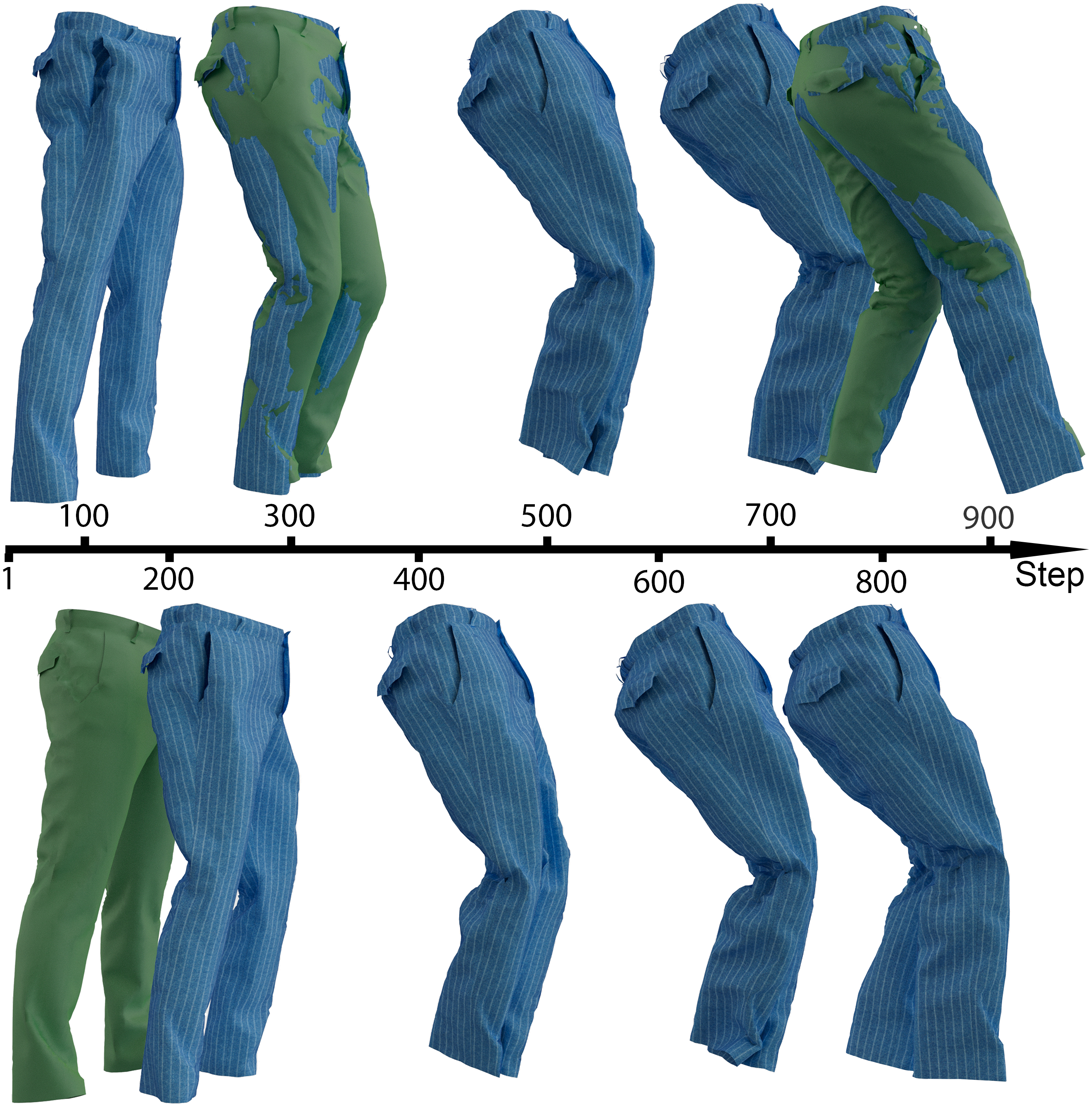}
\caption{\emph{External Force Sequence Optimization.} We find the time-varying force sequence that pushes the garment through the keyframes (green). Over 21 million DoFs are being optimized for.}
        \label{fig:forceOptimizationSimAndKey}
\end{figure}

\subsection{Skeleton Pose Optimization}

\begin{figure*}
  \includegraphics[width=0.98\textwidth]{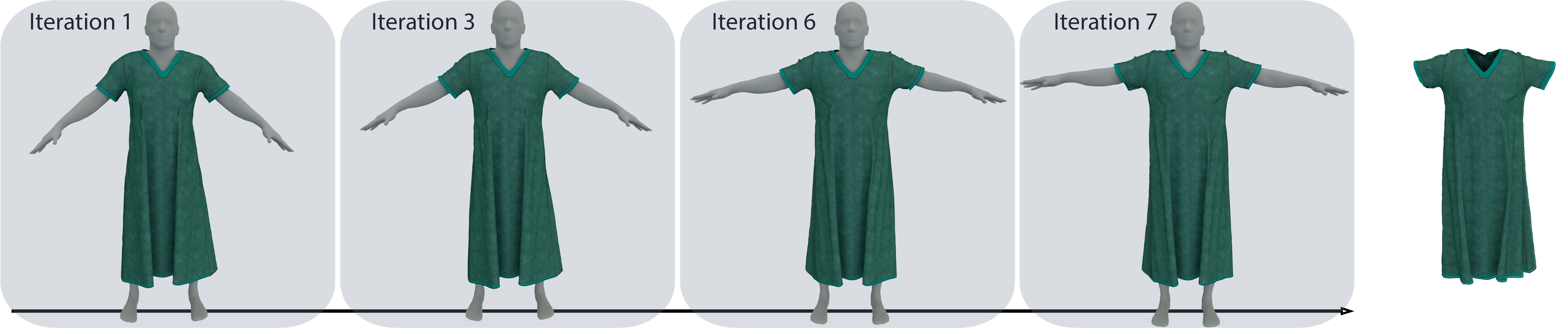}
  \caption{\emph{Skeleton Pose Optimization.} We show how the skeleton pose can be recovered simply by looking at the drape of the garment. From left to right, we visualize different optimization iterations and the target drape is shown on the far right. We optimize over the 3 rotational DoFs of all 114 joints simultaneously.}
  \label{fig:poseEstimation}
\end{figure*}

We can optimize for the skeleton pose that through a linear blend skinning operation defines the body surface positions. We show that we can optimize this pose to produce a draped garment that matches the reference closely. Different iteration results of this experiment are shown in Fig.~\ref{fig:poseEstimation}. 

\subsection{Initial Value Optimization}

\begin{figure*}
  \includegraphics[width=0.98\textwidth]{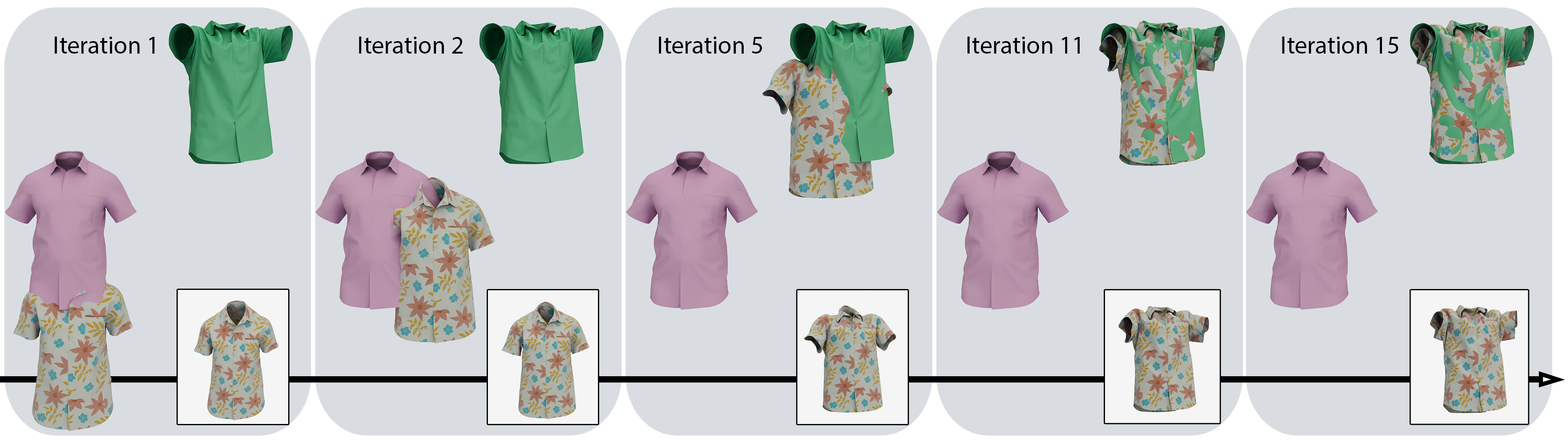}
  \caption{\emph{Initial condition optimization.} We show optimization of the initial velocity per vertex in order to reach a specific pose and location at a specified frame. The pink shirt indicates the starting position and the green is the goal position. Different iterations are visualized. The inset images provide an non-occluded view of the final optimized shape per iteration. Our method effortlessly handles the high DoFs of 197,904 velocity values.}
  \label{fig:initialValueEstimation}
\end{figure*}

We demonstrate optimizing over initial condition parameters. We intend to find the initial velocity per vertex so that at the end of the simulation, the clothing is at a specified position and pose provided by the keyframe. Fig.~\ref{fig:initialValueEstimation} visualizes different iteration results. In the first iteration, the shirt simply falls down due to gravity. In the following iterations, the model iteratively updates the initial velocity prediction such that the shirt coincide closely with the goal shape at the requested frame. This example highlights the scalability of our method by easily estimating all 197,904 velocity values directly. 

\subsection{Keyframe Simulation}

Given a few sparse keyframes of the cloth geometry, we want to optimize the time-varying sequence of forces so that the cloth geometry passes through the provided keyframes at the desired time. We initialize the sequence with forces equal to zero. Initially, the garment falls down under gravity but then converges to a solution where the garment flows through the desired keyframes at the requested time, see Fig.~\ref{fig:forceOptimizationSimAndKey}. This example highlights the capability of our method to optimize for high number of DoFs in an efficient and scalable way. We optimize for 900 simulation steps which results in 21,605,400 control variables being optimized for simultaneously. To the best of our knowledge, this far exceeds prior work. 

\subsection{Performance}

\begin{table}[t]
    \centering
    \resizebox{\columnwidth}{!}{%
    \begin{tabular}{l c c c c c c }
    \toprule
    \multicolumn{1}{c}{\multirow{1}{*}{}} & \multicolumn{1}{c}{\textbf{Simulation}} & \multicolumn{3}{c}{\textbf{Differentiable Simulation}}& \multicolumn{2}{c}{\textbf{Resolution}} \\ \midrule
      & \multirow{2}{*}{\parbox{1.0cm}{\centering Forward}} & \multirow{2}{*}{\parbox{1.0cm}{\centering Forward}}
& \multirow{2}{*}{\parbox{1.3cm}{\centering Matrix Assembly}} & \multirow{2}{*}{\parbox{1.2cm}{\centering CG Solve}} & \multirow{2}{*}{\parbox{1.2cm}{\centering Vertices}} & \multirow{2}{*}{\parbox{1.2cm}{\centering Elements}}    \\ 
     & & & & & & \\ \midrule
    Shirt    & 0.145 & 2.459 & 0.490 & 2.046 & 65,968 & 131,421 \\ 
    T-Shirt  & 0.079 & 0.496 & 0.087 & 0.476 & 14,639 & 29,032 \\ 
    Dress    & 0.056 & 0.354 & 0.064 & 0.282 & 10,422 & 20,685 \\ 
    Pants    & 0.053 & 0.192 & 0.050 & 0.254 & 8002 & 16,170 \\ 
    Swatch   & 0.005 & 0.011 & 0.003 & 0.005 & 441 & 800 \\
    Dino     & 0.039 & 2.425 & 0.006 & 0.007 & 1365 & 4802 \\ \bottomrule
    \end{tabular}
    }
    \caption{\emph{Timings in seconds.} Normal and differentiable simulation timings per step are. Timings are computed as the average per step. Mesh resolutions used in the examples are shown on the right.}
    \label{tab:timingsForward}

\end{table}

The algorithm is implemented in C++ on CPU and we report the timings in Table~\ref{tab:timingsForward}. All experiments are run using an AMD Ryzen Threadripper PRO 3975WX 32-Cores using 20 constraint iterations with a time step of $0.0016$ ms. The timings reported include collision detection and resolving, the constraint solve, and the computation the derivative terms, including definiteness fix. We also report the timing of the non-differentiable version of the solver. Although enabling differentiability adds some expected overhead, the method remains performant. The backward pass consists of matrix assembly from the individual derivative blocks and the linear system solve using Conjugate Gradients.

\subsection{Comparisons To Related Work}
\label{sec:comparisons}

\citet{guo2021inverse} demonstrates body pose and shape optimization from cloth scans. They assume that variations in the time-varying body states only affect the current state of the cloth geometry. This simplification is justified since this would require propagating gradients across the entire simulation resulting in intractable computations. \citet{wojtan2006keyframe} presented a differentiable implicit simulator for resolutions up to 2500 vertices. \citet{liang2019differentiable} demonstrate differentiable cloth simulation on geometries with up to 4096 vertices. \citet{hu2019chainqueen} show optimizations for up to 3000 DoFs. \citet{du2021_diffpd} report gradient computation for resolutions up to nearly 30,000 DoFs. Our examples show that our method is capable of computing gradients and perform optimizations with several orders of magnitude increase in DoFs and increased mesh resolutions. We show up to 65,968 vertices and 21 million DoFs.

\section{Discussion, Limitations, And Future Work}

We present an efficient extension to the position-based simulation model of compliant constraint dynamics to obtain gradients with respect to any parameter through a dynamic simulation. We illustrate the effectiveness of the method with several example applications and show that our method is capable of efficiently computing gradients for high resolutions and high DoFs. 
A limitation of the adjoint method is the need to store intermediate particle states and gradients at every time step. The resulting memory usage scales linearly with the length of the simulation. In practice, this limitation is manageable as our approach allows us to store data for individual steps and retrieve them when needed. Since our contribution does not modify the properties of the forward simulation algorithm, we inherit the same performance advantages but also the limitations such as slow convergence for very stiff constraints. As future work, we would like to use more complex optimization algorithms.  Additionally, we are keen to explore end-to-end optimizations from image data by combining differentiable rendering with differentiable simulation.



\bibliographystyle{ACM-Reference-Format}
\bibliography{main}

\newcommand{\bQ}{\mathbf{Q}}
\newcommand{\bq}{\mathbf{q}}
\newcommand{\bG}{\mathbf{F}}
\newcommand{\bu}{\mathbf{u}}
\newcommand{\bx}{\mathbf{x}}
\newcommand{\bv}{\mathbf{v}}
\newcommand{\bw}{\mathbf{w}}
\newcommand{\bl}{\hat{\mathbf{q}}}
\newcommand{\bmu}{\boldsymbol{\mu}}

\newpage
\section{Supplemental Material}

We review the Adjoint method in Section~\ref{sec:adjoint} and demonstrate how this applies to the simulation of compliant constraint dynamics in section~\ref{sec:diffXPBD}.

\subsection{The Adjoint Method}
\label{sec:adjoint}
Consider the following time dependent optimization problem
$$\min_{\bu} \Phi(\bQ, \bu)$$
\begin{equation}
\begin{aligned}
\textrm{s.t.} \quad \forall n \quad &\bG_n(\bq_{n+1}, \bq_n, \bu) - \bq_{n+1} = \mathbf{0} \\
&\bq_0(\bu) = \mathbf{w}, \\
\end{aligned}
\end{equation}
where $\bq_n = [\bx_n, \bv_n]^\top$, $\bQ = [\bq_0, \bq_1, ..., \bq_N]^{\top}$ is the concatenation of all the discrete state of positions and velocities given some initial some boundary condition $\mathbf{w}$, and $\bG$ is the time integration scheme. Here we choose the commonly used first order implicit time integration scheme, but the derivation can be trivially modified to apply to higher order and explicit methods.
We can write the Lagrangian of this system using the discrete Lagrange multipliers $\bmu$ and $\bl$ as 
$$\mathcal{L} = \bmu^\top (\bq_0 - \mathbf{w}) + \sum_{n=0}^{N-1}\phi_n(\bq_n, \bu) + \bl_n^\top(\bG_n(\bq_{n+1}, \bq_n, \bu) - \bq_{n+1})$$

\begin{equation}
\begin{aligned}
    \frac{d \mathcal{L}}{d \bu} &= \bmu^\top (\frac{d \bq_0}{d \bu} - \frac{d \bw}{d \bu}) + \sum_{n=0}^{N-1} \frac{\partial \phi_n}{\partial \bq_n} \frac{d \bq_n}{d \bu} + \frac{\partial \phi_n}{\partial \bu} +\bl_n^\top \frac{\partial \bG_n}{\partial \bu} \\
    &+ \sum_{n=0}^{N-1} \bl_n^\top (\frac{\partial \bG_n}{\partial \bq_{n+1}} \frac{d \bq_{n+1}}{d \bu} + \frac{\partial \bG_n}{\partial \bq_{n}} \frac{d \bq_{n}}{d \bu} - \frac{\partial \bq_{n+1}}{\partial \bu})
\end{aligned}
\end{equation}

We can rearrange the last term into 
\begin{equation}
    \begin{aligned}
        \sum_{n=0}^{N-1} & \bl_n^\top (\frac{\partial \bG_n}{\partial \bq_{n+1}} \frac{d \bq_{n+1}}{d \bu} + \frac{\partial \bG_n}{\partial \bq_{n}} \frac{d \bq_{n}}{d \bu} - \frac{\partial \bq_{n+1}}{\partial \bu})\\ & = \bl_0^\top \frac{\partial \bG_0}{\partial \bq_0} \frac{d \bq_0}{d \bu} + \bl_{N-1}^\top \frac{\partial \bG_{N-1}}{\partial \bq_N} \frac{d \bq_N}{d \bu} - \bl_{N-1} \frac{d \bG_N}{d \bu} \\ &+ \sum_{n=1}^{N-1} (\bl_{n-1}^\top \frac{\partial \bG_{n-1}}{\partial \bq_n} + \bl_n \frac{\partial \bG_n}{\partial \bq_n} - \bl_{n-1}) \frac{d \bq_n}{d \bu}
    \end{aligned}
\end{equation}

Setting $\bl_{N-1} = 0$ and rearrange the terms
\begin{equation}
    \begin{aligned}
        \frac{d \mathcal{L}}{d \bu} =& \bmu^\top(\frac{d \bq_0}{d \bu} - \frac{d \mathbf{w}}{d \bu}) + \bl_0^\top \frac{\partial \bG_0}{\partial \bq_0} \frac{d \bq_0}{d \bu}\\ &+ \sum_{n=0}^{N-1} \frac{\partial \phi_n}{\partial \bq_n} \frac{d \bq_n}{d \bu} + \frac{\partial \phi_n}{\partial \bu} +\bl_n^\top \frac{\partial \bG_n}{\partial \bu} \\
        &+\sum_{n=1}^{N-1} (\bl_{n-1}^\top \frac{\partial \bG_{n-1}}{\partial \bq_n} + \bl_n \frac{\partial \bG_n}{\partial \bq_n} - \bl_{n-1}) \frac{d \bq_n}{d \bu} \\
        =& (\bmu^\top + \bl_0^\top \frac{\partial \bG_0}{\partial \bq_0} + \frac{\partial \phi_0}{\partial \bq_0})\frac{d \bq_0}{d \bu} - \bmu^\top \frac{d \bw}{d \bu} \\
        &+ \sum_{n=1}^{N-1} (\bl_{n-1}^\top -\frac{\partial \bG_{n-1}}{\partial \bq_n} + \bl_n^\top \frac{\partial \bG_n}{\partial \bq_n} - \bl_{n-1}^\top + \frac{\partial \phi_n}{\partial \bq_n}) \frac{d \bq_n}{d \bu} \\
        &+ \sum_{n=0}^{N-1} \frac{\partial \phi_n}{\partial \bu} + \bl_n^\top \frac{\partial \bq_n}{\partial \bu}
    \end{aligned}
\end{equation}

We can exploit the freedom of the Lagrange multipliers and set
\begin{align}
\bmu^\top + \bl_0^\top \frac{\partial \bG_0}{\partial \bq_0} + \frac{\partial \psi_0}{\partial \bq_0} = 0
\end{align}
and define the adjoint states in the backward difference manner
\begin{equation}
\bl^\top_{n-1} = \bl_{n-1}^\top  \frac{\partial \bG_{n-1}}{\partial \bq_{n}} + \bl_n^\top \frac{\partial \bG_n}{\partial \bq_n} + \frac{\partial \phi_n}{\partial \bq_n}.
\label{eq:adjoint_state}
\end{equation}

The gradient of the Lagrangian can be simplified to the following form
\begin{equation}
    \frac{d \mathcal{L}}{d \bu} = -\bmu^\top \frac{d \bw}{d \bu} + \sum_{n=0}^{N-1} \left( \frac{\partial \phi_n}{\partial \bu} + \bl_n^\top \frac{\partial \bG_n}{\partial \bu}\right).
\end{equation}

A detailed discussion of the continuous adjoint method can be found in the tutorial by ~\citet{bradley2019}, and the discrete adjoint method for forward Euler method by~\citet{betancourt2020discrete}.

\subsection{The Adjoint Method Applied to XPBD}
\label{sec:diffXPBD}
Given the XPBD implicit update rule, $\mathbf{Q} = \mathbf{F}\left( \mathbf{Q}, \mathbf{u}\right)$ takes the form
\begin{equation}
 \begin{bmatrix}
\mathbf{x}_{n+1} \\
\mathbf{v}_{n+1}
\end{bmatrix} = \begin{bmatrix}
    \mathbf{x}_{n} + \Delta \mathbf{x}\left(\mathbf{x}_{n+1} \right) + \Delta t \left( \mathbf{v}_n + \Delta t \mathbf{M}^{-1} \mathbf{f}_{\text{ext}} \right) \\
    \frac{1}{\Delta t} \left( \mathbf{x}_{n+1} - \mathbf{x}_{n} \right)
\end{bmatrix}
\end{equation}
We find the implicit update rule for the adjoint states using~\myeqref{eq:adjoint_state}. For a simulation of $V$ particles we find the adjoint states $\hat{\mathbf{q}}_n$ consisting of adjoint positions $\hat{\mathbf{x}}_n \in \mathbb{R}^{3V}$ and adjoint velocities $\hat{\mathbf{v}}_n \in \mathbb{R}^{3V}$ as
\begin{equation} 
    \hat{\mathbf{q}}_{n-1} = \frac{\partial \mathbf{F}_{n-1}}{\partial \mathbf{q}_n}^\top\hat{\mathbf{q}}_{n-1} + \frac{\partial \mathbf{F}_{n}}{\partial \mathbf{q}_n}^\top\hat{\mathbf{q}}_{n} + \frac{\partial \phi_n}{\partial \bq_n}^\top \\
\end{equation}

\begin{equation}
 \begin{bmatrix}
\hat{\mathbf{x}}_{n-1} \\
\hat{\mathbf{v}}_{n-1}
\end{bmatrix} = \begin{bmatrix}
   \frac{\partial \mathbf{F}_{\mathbf{x}, n-1}}{\partial \mathbf{x}_n} & \frac{\partial \mathbf{F}_{\mathbf{x}, n-1}}{\partial \mathbf{v}_n} \\
    \frac{\partial \mathbf{F}_{\mathbf{v}, n-1}}{\partial \mathbf{x}_n} & \frac{\partial \mathbf{F}_{\mathbf{v}, n-1}}{\partial \mathbf{v}_n}
\end{bmatrix}^\top \begin{bmatrix}
\hat{\mathbf{x}}_{n-1} \\
\hat{\mathbf{v}}_{n-1}
\end{bmatrix} + \begin{bmatrix}
   \frac{\partial \mathbf{F}_{\mathbf{x}, n}}{\partial \mathbf{x}_n} & \frac{\partial \mathbf{F}_{\mathbf{x}, n}}{\partial \mathbf{v}_n} \\
    \frac{\partial \mathbf{F}_{\mathbf{v}, n}}{\partial \mathbf{x}_n} & \frac{\partial \mathbf{F}_{\mathbf{v}, n}}{\partial \mathbf{v}_n}
\end{bmatrix}^\top \begin{bmatrix}
\hat{\mathbf{x}}_{n} \\
\hat{\mathbf{v}}_{n}
\end{bmatrix} + \begin{bmatrix}
    \frac{\partial \phi}{\partial \mathbf{x}_n} \\
    \frac{\partial \phi}{\partial \mathbf{v}_n}
\end{bmatrix}^\top
\end{equation}
Computing and substituting the partial derivative terms relating to $\mathbf{F}$, we find
\begin{equation}
\begin{aligned}
    \hat{\mathbf{x}}_{n} 
    &= \hat{\mathbf{x}}_{n+1} + \left(\frac{\partial \Delta \mathbf{x}}{\partial \mathbf{x}}
    + \Delta t^2 \mathbf{M}^{-1} \frac{\partial \mathbf{f}_{\text{ext}}}{\partial \mathbf{x}} \right)^\top \hat{\mathbf{x}}_{n}
    + \frac{\hat{\mathbf{v}}_n}{\Delta t}
    - \frac{\hat{\mathbf{v}}_{n+1}}{\Delta t}
    + \frac{\partial \phi}{\partial \mathbf{x}}^\top \\
    \hat{\mathbf{v}}_n 
    &= \left( \frac{\partial \Delta \mathbf{x}}{\partial \mathbf{v}} + \Delta t^2 \mathbf{M}^{-1} \frac{\partial \mathbf{f}_{\text{ext}}}{\partial \mathbf{v}} \right)^\top \hat{\mathbf{x}}_{n} 
    + \Delta t \hat{\mathbf{x}}_{n+1} + \frac{\partial \phi}{\partial \mathbf{v}}^\top
\end{aligned}
\label{eq:rawAdjoint}
\end{equation}







\end{document}